# Guided resonances in photonic quasicrystals

Armando Ricciardi,[1] Ilaria Gallina,[2,3] Stefania Campopiano,[1] Giuseppe Castaldi,[2] Marco Pisco,[4] Vincenzo Galdi,[2,*] and Andrea Cusano[4]

[1] *Dept. for Technologies, University of Naples "Parthenope," Centro Direzionale, Isola C4, I-80143, Naples, Italy*
[2] *Waves Group, Dept. of Engineering, University of Sannio, Corso Garibaldi 107, I-82100 Benevento, Italy*
[3] *Dept. of Environmental Engineering and Physics, University of Basilicata, Via dell'Ateneo Lucano 10, I-85100 Potenza, Italy*
[4] *Optoelectronic Division, Dept. of Engineering, University of Sannio, Corso Garibaldi 107, I-82100 Benevento, Italy*
[*] *Corresponding author: vgaldi@unisannio.it*

**Abstract:** In this paper, we report on the first evidence of guided resonances (GRs) in *aperiodically-ordered* photonic crystals, tied to the concept of "quasicrystals" in solid-state physics. Via a full-wave numerical study of the transmittance response and the modal structure of a photonic quasicrystal (PQC) slab based on a representative aperiodic geometry (Ammann-Beenker octagonal tiling), we demonstrate the possibility of exciting GR modes, and highlight similarities and differences with the periodic case. In particular, we show that, as for the periodic case, GRs arise from the coupling of the incident plane-wave with *degenerate* modes of the PQC slab that exhibit a matching symmetry in the spatial distribution, and can still be parameterized via a Fano-like model. Besides the phenomenological implications, our results may provide new degrees of freedom in the engineering of GRs, and pave the way for new developments and applications.

## 1. Introduction

In recent years, defect and bandgap engineering of photonic crystal (PC) slabs have been the subject of many studies aimed at designing micro- and nano-sized optical devices for integrated photonic circuits [1]. Starting from waveguides and cavities, PC slabs (in the form, e.g., of periodic arrangements of dielectric rods or air holes in a host medium) have been proposed as building block for broader applications, including splitters [2], add-drop filters [3,4], modulators [5], slow light [6] and non-channel waveguides [7], just to mention a few.

For applications that require the light-flow control at a wavelength scale, PC slabs made of air hole arrays in a host dielectric medium support guided modes that are completely confined in the slab and cannot couple to any external radiation because of total internal reflection (which results from the refractive index contrast between the slab and the surrounding medium) [8,9]. In addition to guided modes, PC slabs also support *leaky* modes, with finite lifetimes inside the slab, that can couple to the continuum of free-space modes. These latter modes are known as "guided resonances" (GRs) [10], since their electromagnetic power is strongly confined within the slab but, at the same time, they can also couple to the external radiation. In a band diagram, guided modes and GRs are represented below and above, respectively, the light line.

In PC slabs, GRs can be excited under normally incident (with respect to the crystal-periodicity plane) plane-wave illumination [11]. In this situation, the slab transmission/reflection spectra exhibit Fano-like resonance line shapes [12] superimposed on a smoothly varying background that results from the Fabry-Perot effect associated with the light interaction with an effectively-homogeneous dielectric slab. The sharp and asymmetrical resonant features in the spectra are due to the interference between the directly transmitted/reflected wave and those originating from the excited GRs. The GR frequencies and lifetimes depend on the geometric and physical parameters of the PC structure, as well as on the direction and polarization of the incident wave [13].

The intriguing spectral properties of GRs have recently motivated several studies (both numerical and experimental) addressing their characterization and potential applications. From the application viewpoint, their use has been proposed to design micromechanically tunable optical filters [14,15] and polarization-dependent mirrors [13]. Moreover, thanks to their strong sensitivity (frequency shift) to refractive index changes in the surrounding environment, applications have been suggested to biological sensors [16], and also to displacement sensors by using two PC slabs coupled in the near-field regime [17]. From the phenomenological viewpoint, GRs have also been observed in asymmetrical (not free-standing, i.e., not suspended in air) PC slabs on high-index substrates [18]. Moreover, thorough parametric studies have been carried out, both numerically and experimentally (using time-domain terahertz technology) [19]. It is interesting to note that, at variance with the case of in-plane propagation, for the out-of-plane transmission spectra of interest in the GR framework, narrow windows open up without the need of structural defects. Furthermore, it should be emphasized that GRs are not directly related to bandgap-type phenomena [20], and therefore a relatively low refractive index contrast of the PC slab is sufficient for their excitation.

So far, to the best of our knowledge, GRs have been observed only in *periodic* (square and hexagonal) PCs, and it has been argued that they have a strong cooperative nature, with a high degree of spatial coherence reaching out many lattice points. Although it is known that they are intrinsically robust with respect to hole size imperfections and interface roughness in the PC slab (thereby resulting in a higher yield for a given fabrication process) [21], it has also been demonstrated that structural *random* disorder in the hole positions strongly broadens the resonant line shapes in the transmission spectrum, up to their complete disappearance [21]. It appears therefore suggestive, from the phenomenological viewpoint, to explore to what extent the underlying periodicity assumption can be relaxed, by considering intermediate forms of aperiodic (dis)order that lie between the two extrema (perfect periodicity and random disorder). In fact, inspired by the discovery in solid-state physics of "quasicrystals" [22,23] and by the theory of "aperiodic tilings" [24], it has been shown in other PC contexts that periodicity is *not essential* to achieve anomalous (e.g., bandgap, negative refraction and "superfocusing") effects [25,26]. This has led to a growing interest in the study of aperiodically-ordered "photonic quasicrystals" (PQCs) in many fields and applications (see, e.g., [24] for a nice introduction to this subject, and [25,26] for recent reviews). Against this background, in this paper, we investigate the excitation of GRs in PQC slabs based on the *octagonal* (Amman-Beenker, quasiperiodic) square-rhombus tiling [24]. Our investigation involves the three-dimensional (3-D) full-wave study of the transmittance response of the slab, as well as the band-structure and modal analysis for identification of the resonant modes, and shows for the first time that GRs can indeed be excited in PQC slabs, highlighting similarities and differences with respect to the periodic PC case.

Accordingly, the rest of the paper is organized as follows. In Sec. 2, we illustrate the problem geometry, together with the computational tools and observables utilized throughout. In Sec. 3, we present and discuss some representative results concerning the transmittance response (and associated Fano-like parameterization) as well as the band-structure and modal analysis. Finally, in Sec. 4, we provide some concluding remarks and hints for future research.

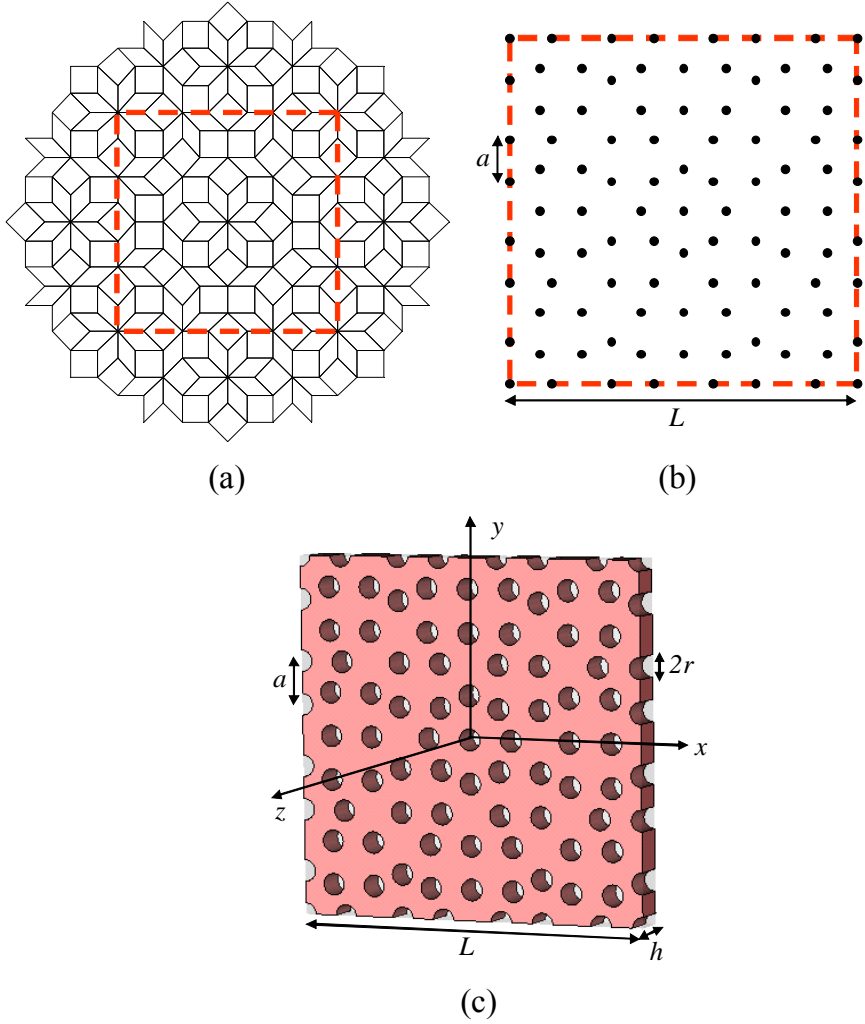

Fig. 1. Problem geometry. (a) A portion of the octagonal (Ammann-Beenker, square-rhombus) tiling, from which the supercell (red dashed square) is cut. (b) Detail of the tiling vertices falling inside the supercell, with indication of the lattice constant $a$ and total size $L$. (c) Final result, in the associated 3-D Cartesian reference system, obtained by placing air holes of radius $r$ in a silicon slab of thickness $h$ at the tiling-vertice positions.

## 2. Problem geometry, observables, and methods

In the study of PQCs, the lack of spatial periodicity prevents straightforward application of well-established theoretical and computational (e.g., Bloch-type) tools and concepts (see, e.g., [27] for some interesting extensions to the quasiperiodic case). Nevertheless, the basic phenomenologies can still be understood by studying a sufficiently large portion of the aperiodic lattice, artificially terminated by suitable periodic boundary conditions. Our investigation relies on this approach, also known as "supercell" approximation, which has already found successful application in many PQC studies [26].

The generation of the PQC slab supercell of interest is illustrated in Fig. 1. We consider an aperiodic lattice based on the (quasiperiodic, eightfold-symmetric) octagonal Ammann-Beenker tiling made of square and rhombus tiles [24]. From the tiling sample displayed in Fig. 1a (generated via standard cut-and-projection method [28]), we cut a sufficiently large square portion (red dashed square), and consider the tiling vertices falling therein. This is more clearly illustrated in Fig. 1b, with indication of the lattice constant $a$ (square/rhombus tile sidelength) and the total sidelength $L$. The final step consists of placing at the tile vertex locations circular air holes (of radius $r$) in a dielectric (silicon) slab of thickness $h$ and refractive index $n=3.418$ immersed in air, as illustrated by the 3-D view in Fig. 1c. In all the simulations below, we assume a hole radius $r = 0.25a$, a slab thickness $h = 0.75a$, and a supercell size $L = (4 + 3\sqrt{2})a$. This results in a PQC-slab supercell composed of 97 holes (or fractions of them), with an air/dielectric fraction $\xi \approx 23.7\%$, and with electric size (in the frequency range of interest) that is compatible with our current computational resources (see also the discussion in Sec. 3.3).

As in [10,19], we study the transmittance response, for time-harmonic $(\exp(-i\omega t))$ excitation, of the PQC-slab supercell in Fig. 1c. In our simulations, we use a commercial software package (CST MICROWAVE STUDIO [29]) based on the finite-integration technique (FIT).

Besides the transmittance spectrum, in order to properly identify the GR modes, we also study the band-structure and corresponding modal field distributions associated with the supercell. In this framework, we utilize a commercial software package (RSOFT BandSOLVE™ [30]) based on a 3-D (vector) plane-wave expansion method.

## 3. Representative results

*3.1 Transmittance response*

In order to minimize the computational burden, it is expedient to take advantage of the supercell mirror symmetries around the horizontal and vertical axes passing through its center (see Fig. 1). In this framework, it is possible reduce the supercell to only one quarter, terminated, e.g., with two horizontal perfectly-electric-conducting (PEC) walls and two vertical perfectly-magnetic-conducting (PMC) walls. Such a "waveguide simulator" supports as the fundamental mode a transverse-electromagnetic (TEM) nondispersive field, which mimics the desired plane-wave excitation propagating along $z$ (i.e., normally-incident on the slab) with $y$-polarized (vertical) electric field. It should be noted, however, that the above equivalence can be reasonably assumed only in the single-mode range of the waveguide simulator, bounded by the cut-off frequency of the first transverse-electric higher-order mode. Figure 2a shows, for the PQC slab under normal-incidence plane-wave excitation, the full-wave-computed transmittance $(|T|^2)$ as a function of the normalized frequency $\nu = \omega a/(2\pi c)$ (with $c$ being the speed of light in vacuum), with frequency-step $\Delta \nu = 10^{-6}$ (black solid curve). The reduced (upper-right quarter) supercell and terminations considered are shown in the inset, and the frequency range is restricted to the single-mode range ($\nu < a/L \approx 0.12$) of the waveguide simulator, so as to ensure that only the fundamental (TEM, plane-wave like) mode is propagating. In the FIT simulations, air layers of thickness $40h$ are assumed at each side of the slab along the $z$ axis, and the entire structure is discretized using a hexahedral mesh of at least 25 lines per wavelength; in the frequency range of interest, this roughly corresponds to 1.38 million mesh cells [29]. Two sharp peaks are clearly visible, and magnified in Figs. 2b and 2c.

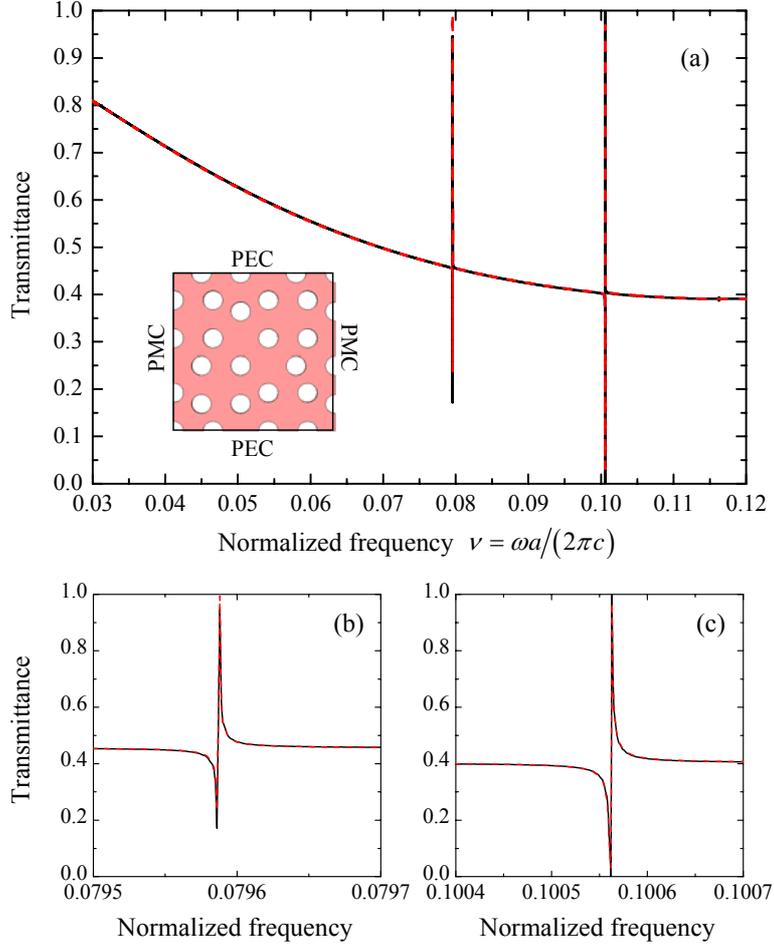

Fig. 2. (a) Full-wave-computed (black solid) and Fano-fitted (red dashed, cf. (1)-(4)) transmittance spectrum (for normal plane-wave incidence, with *y*-polarized electric field) of the PQC slab (reduced supercell and terminations shown in the inset), with $r=0.25a$ (i.e., air/dielectric fraction $\xi \approx 23.7\%$) and $h=0.75a$. (b), (c) Magnified details around the two GRs.

Paralleling [10], the overall transmission spectrum can be parameterized as a superposition of two different responses: *i)* Fabry-Perot smooth oscillations, predominant at lower frequencies and attributable to the finite slab thickness, and *ii)* PQC GR modes, consisting of Fano-like resonant line shapes with widths representing the lifetimes of the modes, viz.,

$$T(\nu) = T_d(\nu) - \sum_k \frac{T_d(\nu) \pm R_d(\nu)}{1 + i\Omega_k}. \tag{1}$$

In (1), $R_d$ and $T_d$ denote the direct reflection and transmission coefficients (representative of the background of the spectra), and $\Omega_k = (\nu - \nu_{0k})/\gamma_k$, with $\nu_{0k}$ and $\gamma_k$ being the normalized central frequency and linewidth, respectively, of the *k*-th GR mode. The plus and minus signs correspond to GR modes that exhibit even (transverse-electric-like) and odd (transverse-magnetic-like) symmetry, respectively, with respect to the mirror plane (*x-y*) parallel to the

slab. Following [10], the behavior of the Fabry-Perot oscillations can be parameterized as follows:

$$R_d(\nu) = \frac{i\frac{(1-\bar{n}^2)}{2\bar{n}}\sin\left(\frac{\bar{n}\nu h}{a}\right)}{\cos\left(\frac{\bar{n}\nu h}{a}\right)-i\frac{(1+\bar{n}^2)}{2\bar{n}}\sin\left(\frac{\bar{n}\nu h}{a}\right)}, \quad T_d(\nu) = \frac{1}{\cos\left(\frac{\bar{n}\nu h}{a}\right)-i\frac{(1+\bar{n}^2)}{2\bar{n}}\sin\left(\frac{\bar{n}\nu h}{a}\right)}, \quad (2)$$

where $\bar{n}$ is a fitting parameter which can be interpreted as the effective refractive index of the PQC slab. Accordingly, besides the full-wave-computed response, Fig. 2 also shows the Fano-fit (red dashed curve) obtained via (1) and (2). As in [10], we first found the frequency-dependent effective refractive index $\bar{n}$ in (2) by assuming a quadratic dispersion model and fitting the background spectrum within the frequency range of interest, which yielded

$$\bar{n}(\nu) = -98.361\nu^2 + 1.534\nu + 6.699. \quad (3)$$

Next, we estimated the (normalized) central frequency $\nu_{0k}$ and linewidth $\gamma_k$ associated with the two peaks ($k$=1,2) via local fits to the Fano-type resonance model

$$T_{GR_k}(\nu) = -\frac{T_d(\nu) \pm R_d(\nu)}{1+i\Omega_k}. \quad (4)$$

The estimated resonance parameters are reported in Table 1 (second and third columns). As it can be observed from Fig. 2, for the above choice of parameters, the Fano-type model agrees very well with the full-wave-computed transmission spectrum (even in the magnified scale of Figs. 2b and 2c), thereby providing strong evidence that the observed peaks are indeed attributable to GRs.

Table 1. Estimated (via Fano-fit) parameters of the first two GRs for a PQC slab (geometry and parameters as in Fig. 2) and a reference periodic (square) PC slab with same thickness, lattice constant, and air/dielectric fraction.

| GR | PQC | | PC (square) | |
|---|---|---|---|---|
| | $\nu_{0k}$ | $\gamma_k$ | $\nu_{0k}$ | $\gamma_k$ |
| $k=1$ | 0.0796 | $2.55 \cdot 10^{-7}$ | 0.3728 | $1.93 \cdot 10^{-4}$ |
| $k=2$ | 0.1006 | $6.27 \cdot 10^{-7}$ | 0.3926 | $3.34 \cdot 10^{-3}$ |

Note that both resonances turned out to be even (plus sign in (4)). Similarly to what observed in [20], the lifetimes of GRs associated with odd modes are larger than those associated with even modes. Therefore, the waves originating from the odd GRs exhibit a long decay time and thus a higher quality factor. As a result, a finer frequency sampling is required to allow the observation of odd GRs in the transmittance spectrum; in addition, a larger computational domain (in terms of air layers in the slab plane, cf. Sec. 3.3 below) could be considered in order to lower their quality factor.

It should be highlighted that Eq. (1) is strictly valid for a maximum of two GRs with opposite parities. Nevertheless, in our example involving two even resonances (as those shown in [10]), it still provides a good fit because the overlap is negligible (i.e., the GR frequency separation is large compared to their linewidths). The reader is referred to [31] for a more general model.

As a reference, in Table 1 (fourth and fifth columns), we also report the GR parameters (also obtained via Fano-fitting the full-wave-computed transmittance spectrum) of a periodic (square) PC slab with same thickness, lattice constant, and air/dielectric fraction. From the comparison, it is interesting to note that the PQC GRs are characterized by moderately lower (nearly a factor four) central frequencies and remarkably smaller (nearly three orders of magnitude) linewidths. In view of the underlying supercell approximation, however, these features are not necessarily representative of *general* aperiodic-order-induced phenomenologies, and deserve further investigations (see also the discussion in Sec. 3.3).

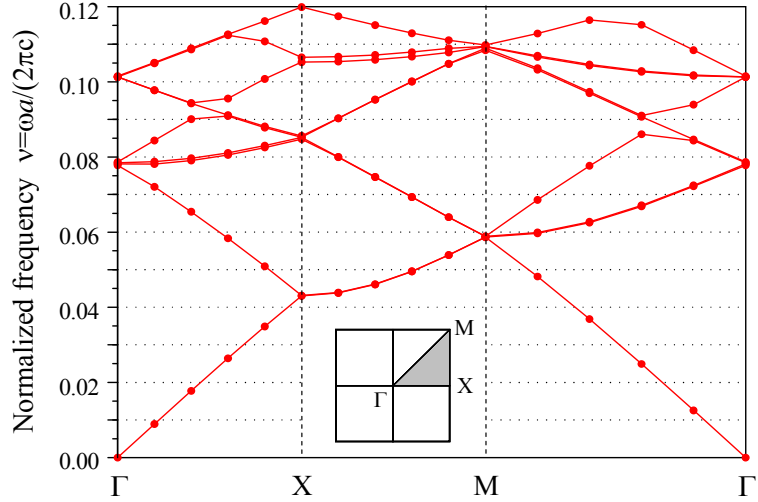

Fig. 3. Band-structure (even modes) of the PQC slab (supercell as in Fig. 1c and parameters as in Fig. 2) along the high symmetry directions of the irreducible Brillouin zone (shown in the inset).

*3.2 Band-structure and modal analysis*

As anticipated, in order to identify the modes responsible for the resonance line shapes in the transmittance response of Fig. 2, we calculated the band diagram of the PQC slab, and subsequently studied the electric field distribution of the excited modes. In our simulations, carried out via a commercial software package (RSOFT BandSOLVE ™ [30]) based on a 3-D (vector) plane-wave expansion method, we considered the entire PQC slab supercell in Fig. 1c, sandwiched between two air layers of thickness $4h$, terminated with 3-D periodic boundary conditions. Note that our study below does not consider odd modes, since they are not responsible for the resonant line shapes in the transmittance response of Fig. 2 (see also the discussion in Sec. 3.1).

Figure 3 shows the computed band-structure for even modes. For the case of actual interest (PQC slab illuminated at normal incidence), attention should be focused on the Γ point (zero in-plane wavevector). The two lowest eigenstates at the Γ point were found to be at normalized frequencies $\nu_1=0.0785$ and $\nu_2=0.1013$, in very good agreement with the GR central frequencies observed in the transmittance response (see Fig. 2 and Table 1).

We then analyzed the modal field distributions at both resonant frequencies. Overall, at each frequency we found four modes, whose electric-field *x*- and *y*- components ($E_{xm}$ and $E_{ym}$, respectively, $m=1,...,8$) are shown in Figs. 4 (at $\nu_1$) and 5 (at $\nu_2$). In order to understand which modes are actually able to couple with the normally-incident plane wave, it is necessary to analyze their symmetry properties. We recall that the electric field amplitude of a plane wave propagating along the *z*-axis possesses an even symmetry with respect to both the *x* and *y* axes. Clearly, for a plane wave with *y*-polarized electric field (as in Fig. 2), only the *y* components of the modal fields of Figs. 4 and 5 need to be taken into account in order to ascertain the coupling. By observing the modal field distributions at normalized frequency $\nu_1$, it is immediate to realize that $E_{y3}$ turns out to be the only *y*-component exhibiting even symmetries with respect to both the *x* and *y* axis. Analogously, considering the resonant frequency $\nu_2$, $E_{y6}$ is the only *y*-component that exhibits similar symmetry properties. Hence, a normally incident *y*-polarized plane-wave is able to couple only with the *m*=3 (at $\nu_1$) and *m*=6 (at $\nu_2$) modes. All other modes exhibit mirror symmetries that prevent the coupling.

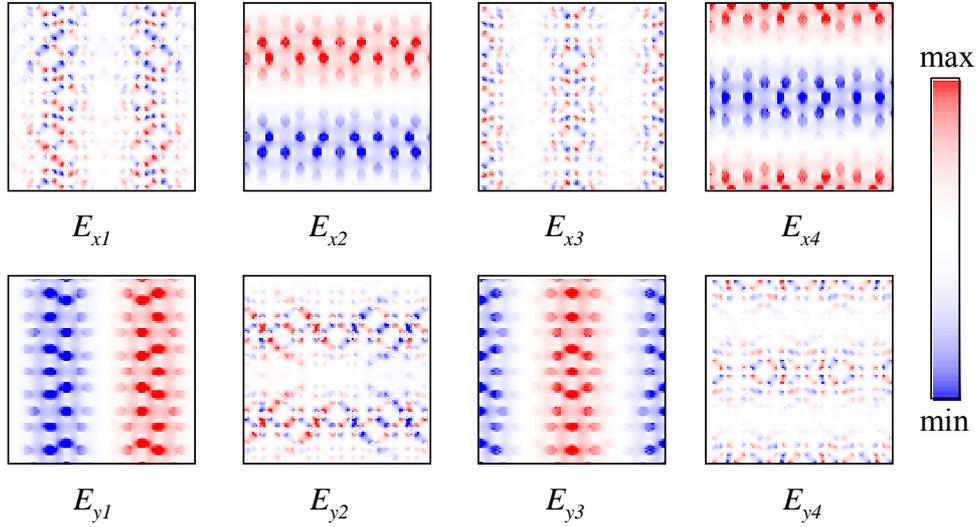

Fig. 4. Electric modal field amplitudes (in the *x-y* plane, at *z*=0) for *x* and *y* components ($E_{xm}$ and $E_{ym}$, respectively), at normalized frequency $\nu_1$=0.0785. Indexes *m*=1,..,4 label the four different modes.

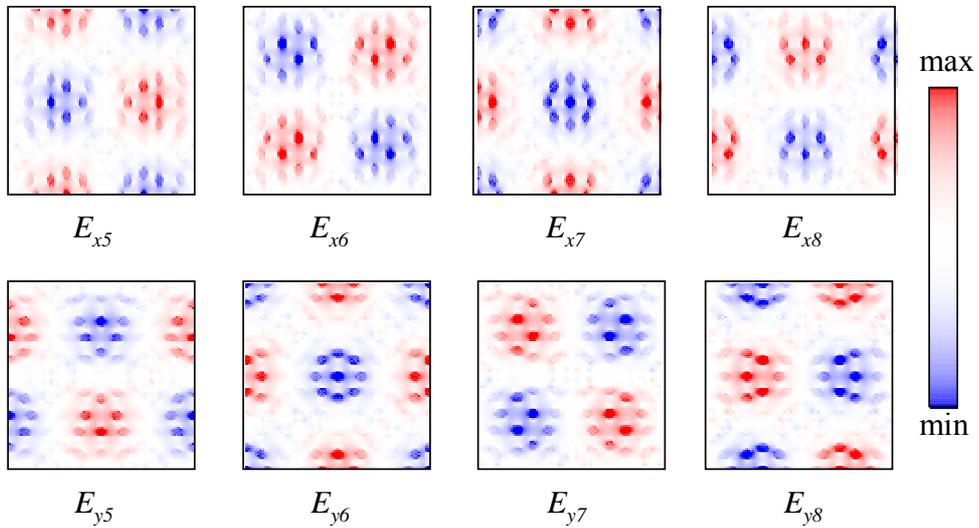

Fig. 5. As in Fig. 4, but at normalized frequency $\nu_2$=0.1013. Indexes *m*=5,..,8 label the four different modes.

Similar arguments can be used to identify the coupled modes for an *x*-polarized plane-wave illumination: In this case, only the *m*=4 (at $\nu_1$) and *m*=7 (at $\nu_2$) modes can be excited.

Moreover, the same results can be obtained if the symmetry considerations are referred to the electric field directions [32,33]. Similar considerations also apply to the case of odd modes.

It is interesting to note that the above coupled modes form a *doubly degenerate* pair: The $m=3$ and $m=4$ modes (at $v_1$), as well as the $m=6$ and $m=7$ modes (at $v_2$), are 90°-rotated versions of each other. This resembles what observed in the periodic PC case, for which group-theory-based studies have demonstrated that the coupling between a normally-incident plane wave and GRs does involve only degenerate modes that exhibits a matching symmetry in the spatial-distribution [34,35].

*3.3 Remarks on the supercell approximation*

The above results, concerning the transmittance response as well as the band-structure and modal analysis, pertain to a supercell size $L = (4+3\sqrt{2})a \approx 8.24a$. As previously mentioned, this choice is related to computational affordability considerations. One might therefore argue that, strictly speaking, the observed features still arise from spatial periodicity (though on a much longer range), and may actually be practically unobservable (in view of their very small linewidth) in a large but finite-size PQC slab. To better understand these aspects, and ascertain which observed features may be reasonably attributed to genuine aperiodic-order-induced phenomenologies, we carried out extensive parametric studies.

First, we gradually increased the supercell size up to the maximum affordable value (taking also into account the waveguide-simulator single-mode constraint), and observed the presence of consistently similar spectral features and modal structures. For instance, Fig. 6 shows the normalized central frequency $v$ and linewidth $\gamma$ pertaining to the higher-frequency GR in Fig. 2, for different values of the supercell size $L$. The central frequency turns out to exhibit a slightly decreasing trend (with increasing the supercell size). While no clear trend can be extracted from the linewitdth behavior, it appears that, at least within the numerical and fitting approximations involved, the GR lineshapes do not progressively narrow down as the supercell size is increased. Note that we verified via modal analysis (see Sec. 3.2) that the GRs involved were all consistently related to a $y$-polarized $m=6$ mode (cf. Fig. 5).

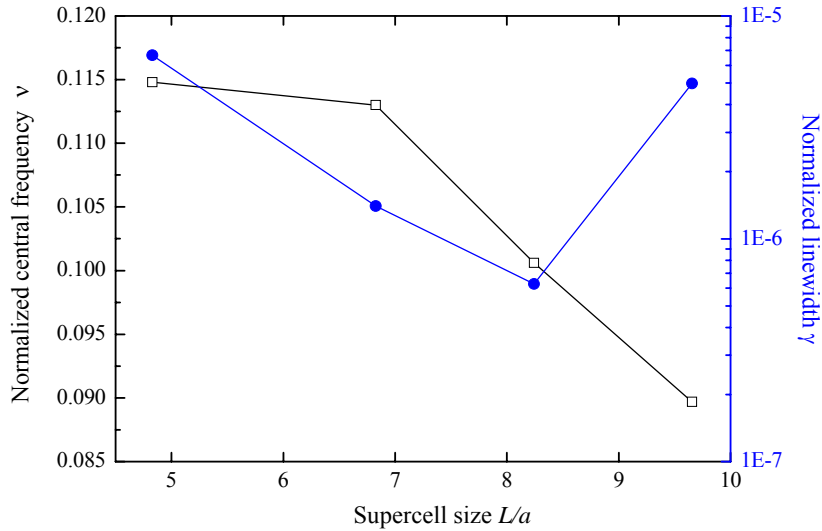

Fig. 6. GR normalized central frequencies (white squares, left axis) and linewidths (blue circles, right axis) for various values of the supercell size.

As a further check, we tried to simulate a *finite-size* PQC slab by considering a *composite* supercell where an air layer of variable thickness $\Delta$ is placed around the usual PQC sample of size $L = \left(4 + 3\sqrt{2}\right)a$ (see the upper-right quarter in the inset of Fig. 7). For increasing values of the air layer thickness, one expects to progressively reduce (also via total internal reflection) the coupling effects with the neighboring periodic replicas, and thus the possible periodicity-induced artifacts. Figure 7 shows the transmittance response pertaining to three representative values of Δ, together with the reference response in the absence of the air layer (Δ=0, cf. Fig. 2); the comparison is restricted to a frequency range around the lowest-frequency GR in order to avoid propagation of higher-order modes in the waveguide simulator. Interestingly, one can always observe a GR-like lineshape, progressively moving from the reference one (for small values of Δ) to lower frequencies (for increasing values of Δ) with a *considerable broadening* of the resonances linewidths (of the order of those observed in the periodic case). Moreover, by comparison with the reference case (no air layer) one also observes a shift towards higher values of the Fabry-Perot background, readily explained by taking into account the direct coupling ensured by the air layer. It is worth noting that, in spite of the intrinsic periodicity of the numerical model, the introduction of air layers of increasing size, with the consequent progressive reduction of the coupling effects among adjacent slabs replicas, does not lead to the disappearance of the GRs.

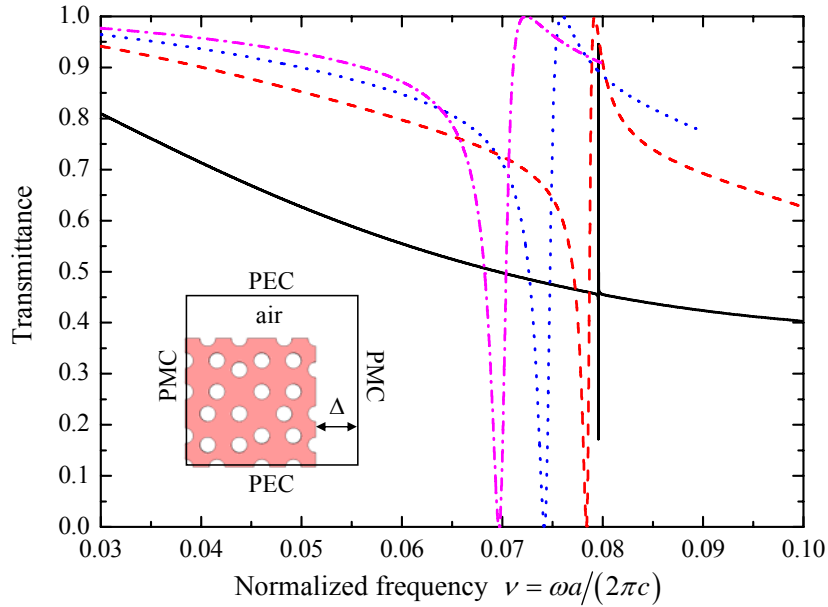

Fig. 7. As in Fig. 2a, but for a composite supercell featuring an air layer of thickness Δ placed around the PQC sample (see inset). Black solid curve: Δ=0 (no air layer); red dashed: Δ=0.88*a*; blue dotted: Δ=1.5*a*; magenta dashed-dotted: Δ=2.13*a*.

In order to further assess the meaningfulness of the above results, we studied the corresponding modal field distributions (cf. Sec. 3.2). Figure 8 shows the resonant-mode field map (*y*-component, computed via RSOFT BandSOLVE™ [30]) for the case of an air layer of thickness Δ=2.13*a* (the largest value compatible with the waveguide-simulator single-mode condition, and corresponding to an inter-replica separation of about one third of a vacuum wavelength). One can observe a field structure clearly resembling the *y*-polarized *m*=3 mode in Fig. 4, pretty well localized in the PQC slab (with a significant decay in the air layer), thereby implying rather *weak* interactions among adjacent periodic replicas.

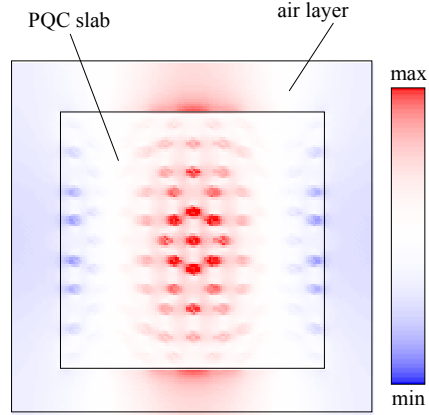

Fig. 8. Electric modal field amplitudes (*y*-component, in the *x-y* plane, at *z*=0) for a composite supercell featuring an air layer of thickness Δ=2.13*a*. Note the resemblance with the *y*-polarized *m*=3 mode in Fig. 4, and the significant field decay in the air layer.

To sum up, the above analysis suggests that the observed GR excitation is not attributable to artifacts generated by the artificial periodic truncation, but is indeed representative of genuine aperiodic-order-induced phenomenologies. The same conclusion cannot safely be drawn for the very small linewidth values observed, which deserve further investigations.

**4. Conclusions**

In this paper, we have presented a 3-D numerical study of the response of PQC slabs (made of aperiodically-ordered arrays of air holes in a dielectric host medium) to normally-incident plane-wave excitation. With reference to a representative aperiodic geometry based on the (quasiperiodic, eightfold-symmetric) Ammann-Beenker octagonal tiling, and via a full-wave study of the transmittance response and of the modal structure, we have shown the possibility (via coupling with degenerate modes of the PQC slab exhibiting proper spatial symmetries) of achieving sharp, asymmetrical Fano-like resonant line shapes in the transmittance response, amenable to the GR modes observed in periodic PC slabs [10].

To the best of our knowledge, this represents the first evidence of GRs in PQC slabs, and, complementing certain results in the topical literature pertaining to *randomly disordered* geometries (see, e.g., [21]), indicates that perfect spatial periodicity is not an essential requirements for their excitation.

Besides the phenomenological implications, our results endow with new perspectives and degrees of freedom in the engineering of GRs, and pave the way for new developments and applications. In this framework, current and future investigations are aimed at a systematic comparative study of the GR mode lifetimes and linewidths in PC and PQC slabs, as well as at the exploitation of the lattice (a)symmetry as a further device in the control of the uncoupled GRs [32]. Also of interest, and currently being pursued, is the experimental validation of the results.

**Acknowledgments**

This work was supported in part by the Italian Ministry of Education and Scientific Research (MIUR) under a PRIN-2006 grant, and in part by the Campania Regional Government via a 2006 grant (L.R. N. 5 - 28.03.2002). The kind assistance of Prof. M. N. Armenise (Polytechnic of Bari, Italy) in the band-structure simulations is gratefully acknowledged.